\documentclass[a4paper]{jpconf}
\usepackage{graphicx}

\pdfoutput=1

\usepackage{mathrsfs}
\DeclareSymbolFontAlphabet{\mathrsfs}{rsfs}
\newcommand{\scri}{\mathrsfs{I}}

\newcommand{\aconf}{\bar\Omega}
\newcommand{\CZ}{Z4c}

\begin{document}
\title{Unconstrained hyperboloidal evolution of black holes in spherical symmetry with GBSSN and Z4c}

\author{Alex Va\~n\'o-Vi\~nuales and Sascha Husa}

\address{Universitat de les Illes Balears and Institut d'Estudis Espacials de Catalunya, Cra. de Valldemossa km.7.5, 07122 Palma de Mallorca, Spain}

\ead{alex.vano@uib.es}

\begin{abstract}
We consider unconstrained evolution schemes for the hyperboloidal initial value problem in numerical relativity as a promising candidate for the optimally efficient numerical treatment of radiating compact objects. Here, spherical symmetry already poses nontrivial problems and constitutes an important first step to regularize the resulting singular PDEs. We evolve the Einstein equations in their generalized BSSN and Z4 formulations coupled to a massless self-gravitating scalar field. Stable numerical evolutions are achieved for black hole initial data, and critically rely on the construction of appropriate gauge conditions.
\end{abstract}

\section{Introduction}

For isolated systems, their energy loss and radiation are in general only well defined at null infinity ($\scri^+$). This is where the observers of astrophysical events are located \cite{Barack:1998bv,Leaver1986,PhysRevD.34.384} and where we ideally want to extract radiation signals from numerical simulations. 
Following the approach of Penrose \cite{PhysRevLett.10.66,Penrose:1965am}, we conformally compactify spacetime: we rescale the physical metric $\tilde g_{\mu\nu}$ by a conformal factor that vanishes at $\scri^+$: 
\begin{equation}\label{rescmetric}
g_{\mu\nu} \equiv\Omega^2\tilde g_{\mu\nu} \ . 
\end{equation}
The equations of motion in terms of the rescaled metric $g_{\mu\nu}$ diverge at $\scri^+$:
\begin{equation}\label{eq:EEconformal}
G_{\mu\nu}[g] = 8\pi\ T_{\mu\nu} -\frac{2}{\Omega}\left(\nabla_\mu\nabla_\nu\Omega-g_{\mu\nu}\nabla^\gamma\nabla_\gamma\Omega\right)-\frac{3}{\Omega^2}g_{\mu\nu}(\nabla_\gamma\Omega)\nabla^\gamma\Omega \ . 
\end{equation}

The character of an initial value formulation depends on how we foliate spacetime, unfortunately all choices which reach future null infinity come with significant technical problems. Characteristic foliations offer many simplifications  \cite{Winicour2009LRR}, but are also prone to develop caustics in the strong field region. Cauchy-characteristic matching, where data on spacelike slices are matched to characteristic slices which reach $\scri^+$ \cite{BishopCCM,Winicour2009LRR,Reisswig:2009us,Reisswig:2009rx,Taylor:2013zia}, poses considerable difficulties to find a stable algorithm for the matching procedure. A smooth way of combining spacelike foliations in the strong field region and also reaching future null infinity is offered by the hyperboloidal initial value approach, pioneered by Friedrich \cite{friedrich1983,lrr-2004-1,Friedrich:2003fq}. Here one evolves along hyperboloidal slices, i.e.~spacelike slices that reach null infinity. In this approach regularizing the singular equations (\ref{eq:EEconformal}) in a way that is not just numerically stable, but also avoids instabilities arising from the continuum equations, has proven to be difficult, 
in particular for hyperbolic free evolution schemes. For the use of elliptic-hyperbolic systems see \cite{Andersson:springer,Rinne:2009qx,Rinne:2013qc}. In this paper we consider the already nontrivial hyperboloidal initial value problem in spherical symmetry. 
 We have recently presented stable evolutions of regular initial data which do not form black holes \cite{Vano-Vinuales:2014koa}, here we use our code to evolve a scalar field interacting with a black hole as a problem to test how our algorithm handles black holes, and we demonstrate that our code can track the expected power law tails. 

\section{Hyperboloidal foliations}

\begin{figure}[h!!!!]
\center
\begin{tabular}{@{}c@{}@{}c@{}}
\mbox{\includegraphics[width=0.5\linewidth]{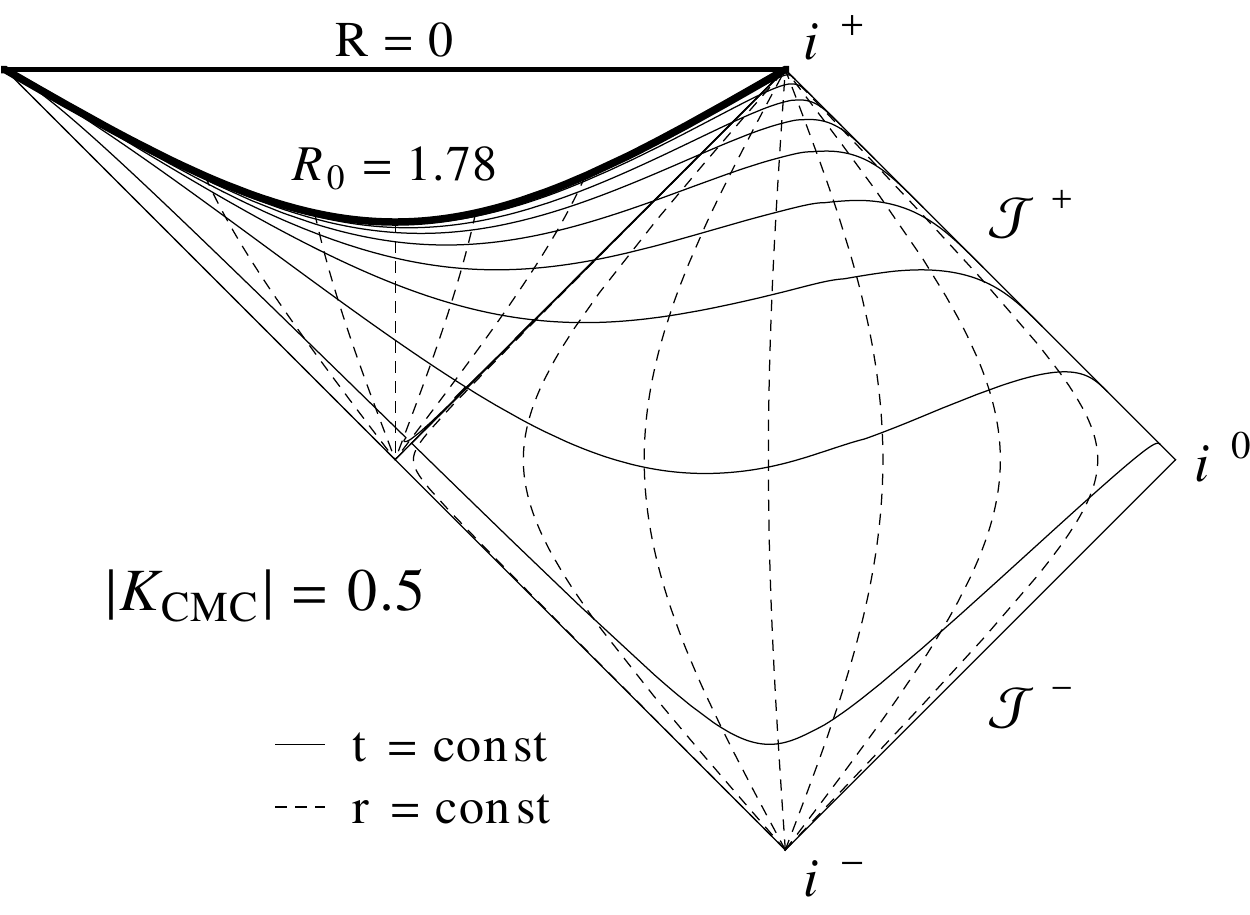}} &
\mbox{\includegraphics[width=0.5\linewidth]{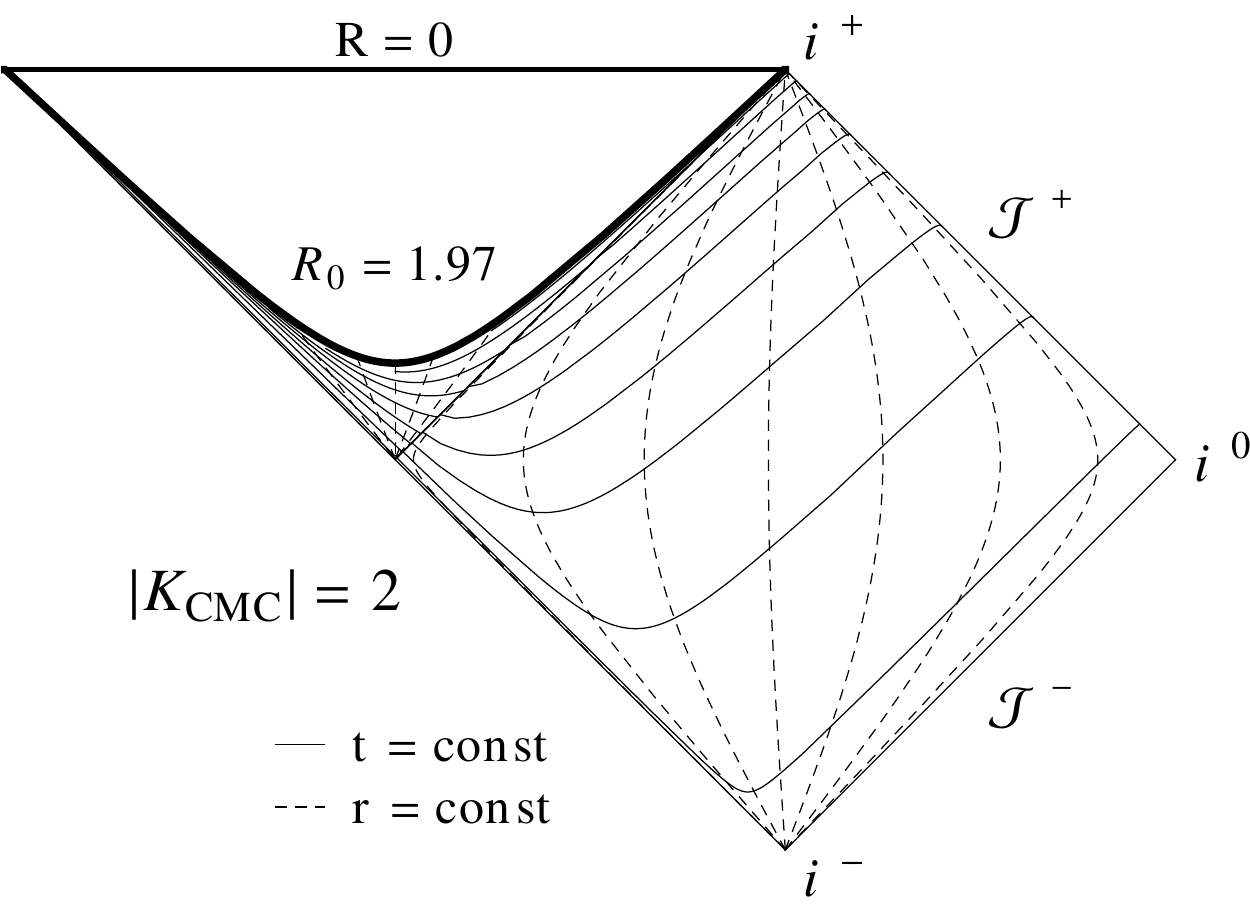}}
\end{tabular}
\caption{Carter-Penrose diagrams showing foliations for two values of $K_{CMC}$. The thick line corresponds to the smallest value of the Schwarzschild radial coordinate, $R_0$, along the slice.}
\label{hypfol}
\end{figure}

Starting with the Schwarzschild metric 
\begin{equation}
 d\tilde s^2 = -A(\tilde r)d\tilde t^2+B(\tilde r)d\tilde r^2+\tilde r^2 d\sigma^2, \quad
  d\sigma^2\equiv d\theta^2+\sin^2\theta d\phi^2,
\end{equation}
where $A(\tilde r) = B(\tilde r)^{-1} = 1-\frac{2M}{\tilde r}$ and the parameter $M$ denotes the mass of the black hole, we first transform the Schwarzschild time coordinate $\tilde t$ to a new time coordinate $t$ whose constant values determine the hyperboloidal slices. We then compactify the radial coordinate and conformally rescale the line element according to \eref{rescmetric}: 
\begin{equation}
t = \tilde t-h(\tilde r), \qquad  \tilde r=\frac{r}{\aconf}  \qquad\textrm{and}\qquad d\bar s^2 = \Omega^2d\tilde s^2,
\end{equation}
where $h(\tilde r)$ is the height function (compare e.g.~\cite{Malec:2003dq}),  $\Omega$ is the conformal factor used to rescale the metric, and $\aconf$ the compactifying factor to rescale the radial coordinate. The resulting line element is
\begin{equation}\label{fsthyp}
d\bar s^2= -A\Omega^2dt^2+\frac{\Omega^2}{\aconf^2}\left[-2Ah'(\aconf-r\aconf')dtdr+(B-Ah'^2)\frac{(\aconf-r\aconf')^2}{\aconf^2}d r^2 + r^2 d\sigma^2\right].
\end{equation}
Here $A$ and $B$ are now functions of $r/\aconf$, and the spatial derivative of the height function is
\begin{equation}
h'(\tilde r)=\frac{dh}{d\tilde r}=-\frac{\frac{K_{CMC}\tilde r^3}{3}+C_{CMC}}{\left(1-\frac{2M}{\tilde r}\right)\sqrt{\left(\frac{K_{CMC}\tilde r^3}{3}+C_{CMC}\right)^2+\left(1-\frac{2M}{\tilde r}\right)\tilde r^4}}. 
\end{equation}
The parameter $K_{CMC}$ is the trace of the extrinsic curvature that labels the Constant-Mean-Curvature (CMC) foliation and $C_{CMC}$ is an integration constant.
For given values of $M$ and $K_{CMC}$, there is only one possible choice for $C_{CMC}$ that will give us trumpet  initial data \cite{Hannam:2006vv}. We also obtain the smallest value of the Schwarzschild radius $\tilde r=R_0$ that the hyperboloidal foliation can reach, and where the trumpet is located at  infinite proper distance from the apparent horizon. Two examples of hyperboloidal CMC foliations of Schwarzschild are shown in figure \ref{hypfol}; the larger the value of $|K_{CMC}|$, the closer the hyperboloidal slices are to characteristic ones. 

For the conformal factor $\Omega$ that rescales the metric we follow Zengino\u{g}lu \cite{Zenginoglu:2007jw,Zenginoglu:2008wc,Zenginoglu:2008pw,Zenginoglu:2008uc} and use a conformal factor fixed in time, whose expression is
\begin{equation}\label{omeg}
\Omega=\frac{r_{\!\!\scri}^2-r^2}{6\, r_{\!\!\scri}}|K_{CMC}| \ , 
\end{equation}
where $r_{\!\!\scri}$ is the position of null infinity, which in the following we will set to unity. 

\begin{figure}[htbp!!]\label{omegas}
\center
	\includegraphics[width=0.6\linewidth]{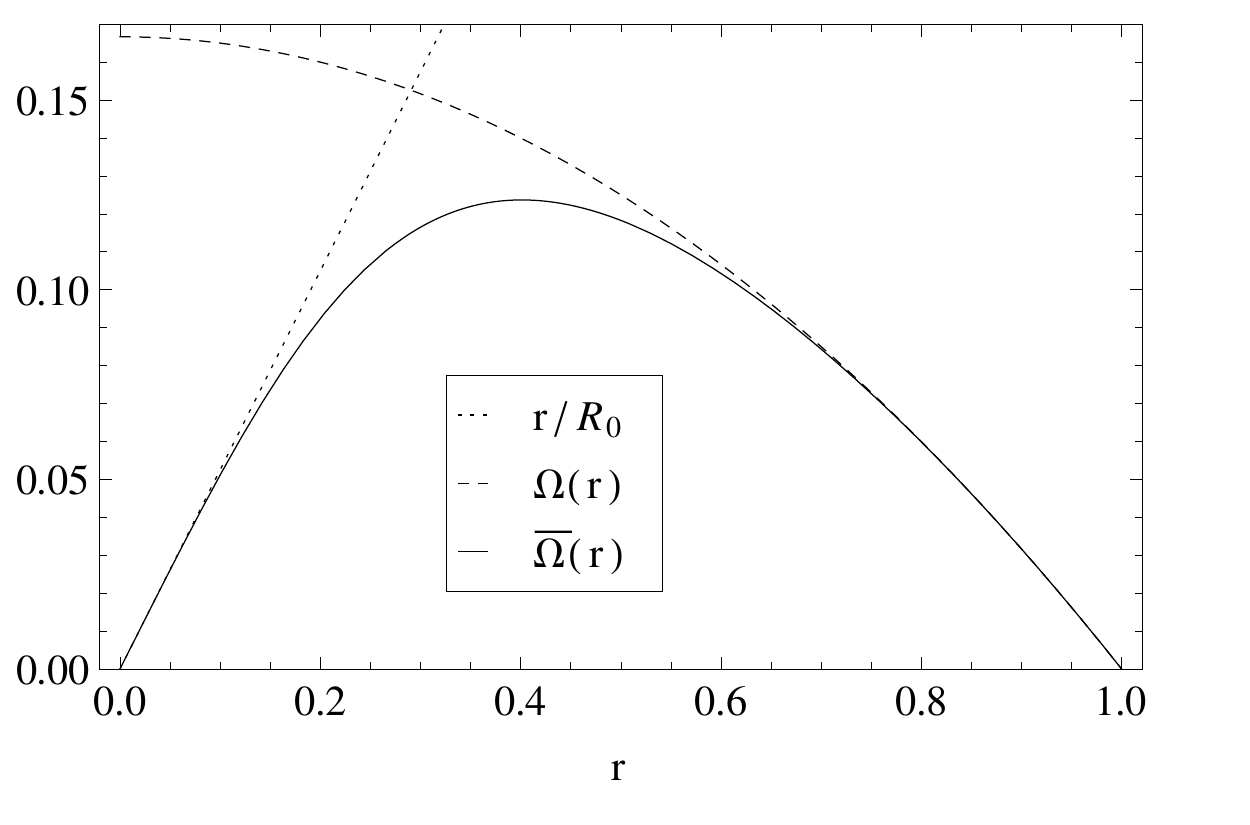}
	\vspace{-2ex}
\caption{Factors $\Omega$ and $\aconf$ for $M=1$ and $K_{CMC}=-1$.}
\end{figure}

The value of $\aconf$ is determined numerically by imposing a conformally flat spatial metric. The behaviour of
 $\aconf$ near $r=0$ is linear in the compactified radius $r$, with a slope inversely proportional to the minimal Schwarzschild radius, $R_0$. In figure \ref{omegas} we show $\Omega$, $\aconf$ and its approximation at the origin as a function of $r$. Note that at $r=r_{\!\!\scri}$, $\Omega$ equals $\aconf$, as at null infinity the spacetime is asymptotically flat. 

\section{Conformally compactified equations}

We use either the generalized BSSN formulation \cite{PhysRevD.52.5428,Baumgarte:1998te,Brown:2007nt} or a similar conformal version of the Z4 formulation \cite{bona-2003-67,Alic:2011gg,Sanchis-Gual:2014nha}, the \CZ{} equations \cite{Bernuzzi:2009ex,Weyhausen:2011cg}, in their spherically symmetric reduction. The actual equations used in the simulations can be found in Appendix C of \cite{Vano-Vinuales:2014koa}. 
The massless scalar field satisfies the wave equation in the physical spacetime, and with respect to the unphysical rescaled metric it satisfies
\begin{equation}
g^{\mu\nu}\nabla_\mu\nabla_\nu\Phi-2g^{\mu\nu}\nabla_\mu\Phi\frac{\nabla_\nu\Omega}{\Omega}=0. 
\end{equation}

Following Zengino\u{g}lu \cite{Zenginoglu:2007it,Zenginoglu:2008pw}, we use a fixed conformal factor $\Omega$ \eref{omeg}, and a scri-fixing gauge where we demand that $\left({\partial}/{\partial t}\right)^a=\alpha n^a+\beta^a$ becomes null at $\scri^+$. 
Since $\scri$ is a null surface, our condition will be satisfied if $\left({\partial}/{\partial t}\right)^a$ is parallel to $\nabla^a\Omega$ at $\scri$, and we obtain $\left.\partial_t\Omega\right|_{\scri}=0$, consistent with the value of $\Omega$ not being evolved in time. 
The time vector $\left({\partial}/{\partial t}\right)^a$ is null at $\scri^+$ if $ \left.-\alpha^2+\beta^a\beta_a\right|_{\scri}=0$, which
in our variables becomes $\left.-\alpha^2+\chi^{-1}\gamma_{rr}{\beta^r}^2\right|_{\scri}=0$. In our simulations this is achieved by making sure that the values of $\alpha$ and $\beta^r$ stay fixed at $\scri^+$ during the evolution. 

The shift can either be fixed throughout the evolution or evolved with the rest of the variables. For black hole initial data we choose the second option. 
As evolution equation we chose a generalized Gamma-driver shift condition: 
\begin{eqnarray}
\dot \beta^r & = & \beta^r{\beta^r}'+\frac{3}{4} \mu B^r + L_0 -\frac{\xi_{\beta^r}}{\Omega}\beta^r , \\ 
\dot B^r & = & \beta^r {B^r}'-\eta B^r +\lambda\left(\dot \Lambda^r-\beta^r{\Lambda^r}'\right)  . 
\end{eqnarray}	
A source function $L_0$ calculated from the flat spacetime values was added, as well as a damping term ($-\beta^r/\Omega$) that makes sure that the value of $\beta^r$ stays fixed at $\scri^+$. The parameters $\lambda$ and $\mu$ have to be chosen carefully, as they determine the eigenspeeds. Appropriate values for our simulations are: $\xi_{\beta^r}=5$, $\lambda=3/4$, $\mu=0.15+4(1-r^2)$ and $\eta=0.1$. 

As slicing condition we choose a generalized Bona-Mass\'o equation that takes the form 
\begin{equation}
\dot \alpha=\beta^r\alpha'-f(\alpha)\left(K-K_0\right)+L_0 ,
\end{equation}
where we have the freedom to choose the two functions $K_0$ and $L_0$. 
The presence of $K_0$ is necessary to prevent the equation from growing exponentially, as the extrinsic curvature $K$ is negative in its stationary solution. 
The source function $L_0$ is calculated from flat spacetime initial data on the hyperboloidal foliation. 
Due to their properties, we are interested in having the 1+log slicing condition around the origin and the harmonic one near $\scri^+$. For this we can make $f(\alpha)$ change smoothly from $2\alpha$ to $\alpha^2$. For the numerical results shown below we have used 
$$\dot \alpha=\beta^r\alpha'-\alpha^2\left(K-K_0\right)+L_0 +(\alpha_0-\alpha),$$
where $\alpha_0$ are the initial data corresponding to the Schwarzschild lapse.

\section{Numerical test: power law tails of the scalar field}

\begin{figure}[htbp!!]\label{convphitails}
\center
	\includegraphics[width=0.74\linewidth]{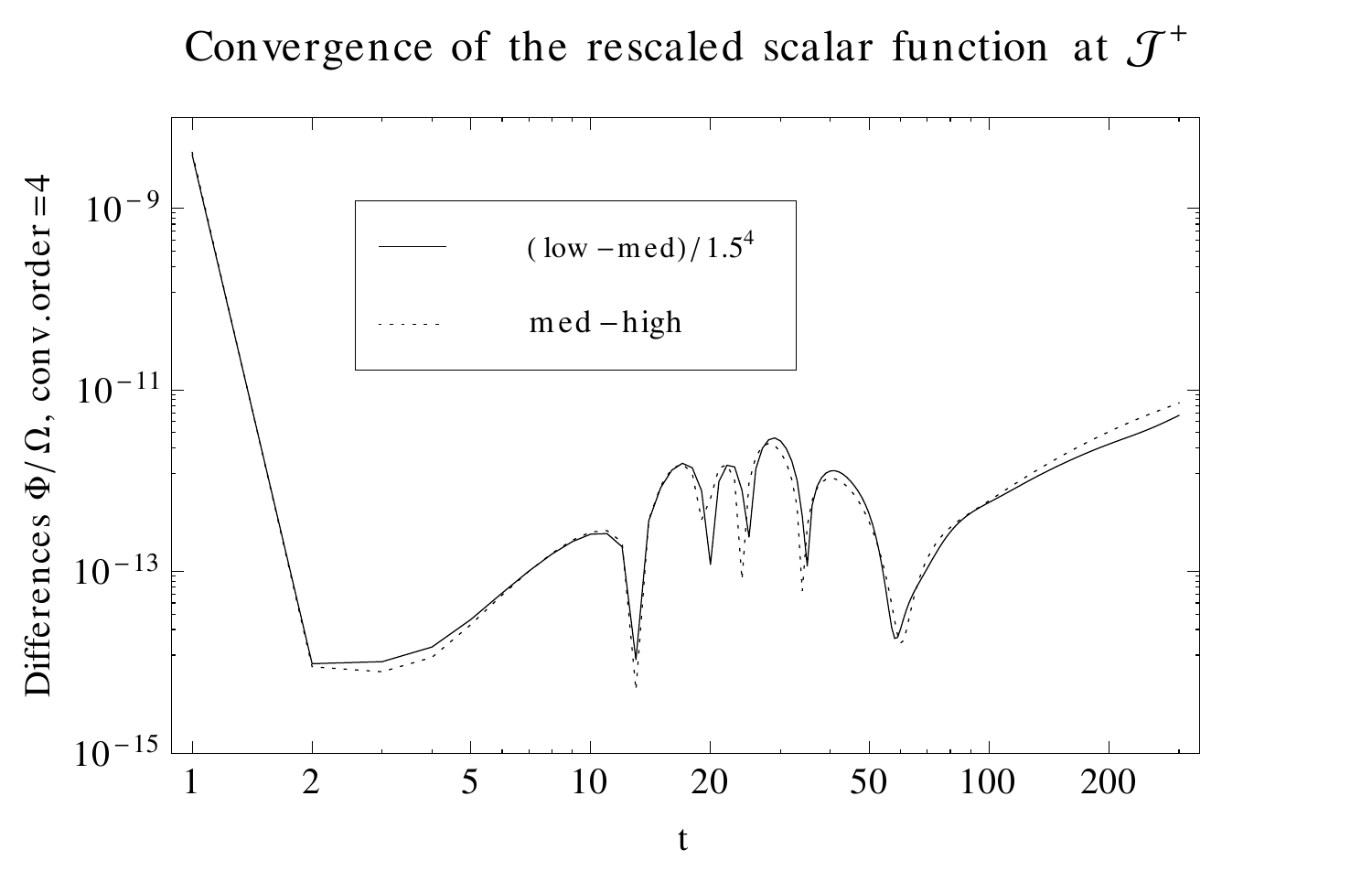}
	\vspace{-2ex}
\caption{Convergence of the rescaled scalar field at $\scri^+$.}
\end{figure}

The equations were implemented in a spherically symmetric code that uses the method of lines with a 4th order Runge-Kutta time integrator and 4th order finite differences. We add Kreiss-Oliger dissipation \cite{kreiss1973methods} and evolve on a staggered grid, which does not include $\scri^+$, however we obtain 4th order convergence for values interpolated to $\scri^+$ as seen in Fig.~\ref{convphitails}.

\begin{figure}[htbp!!]\label{phitails}
\center
	\includegraphics[width=0.84\linewidth]{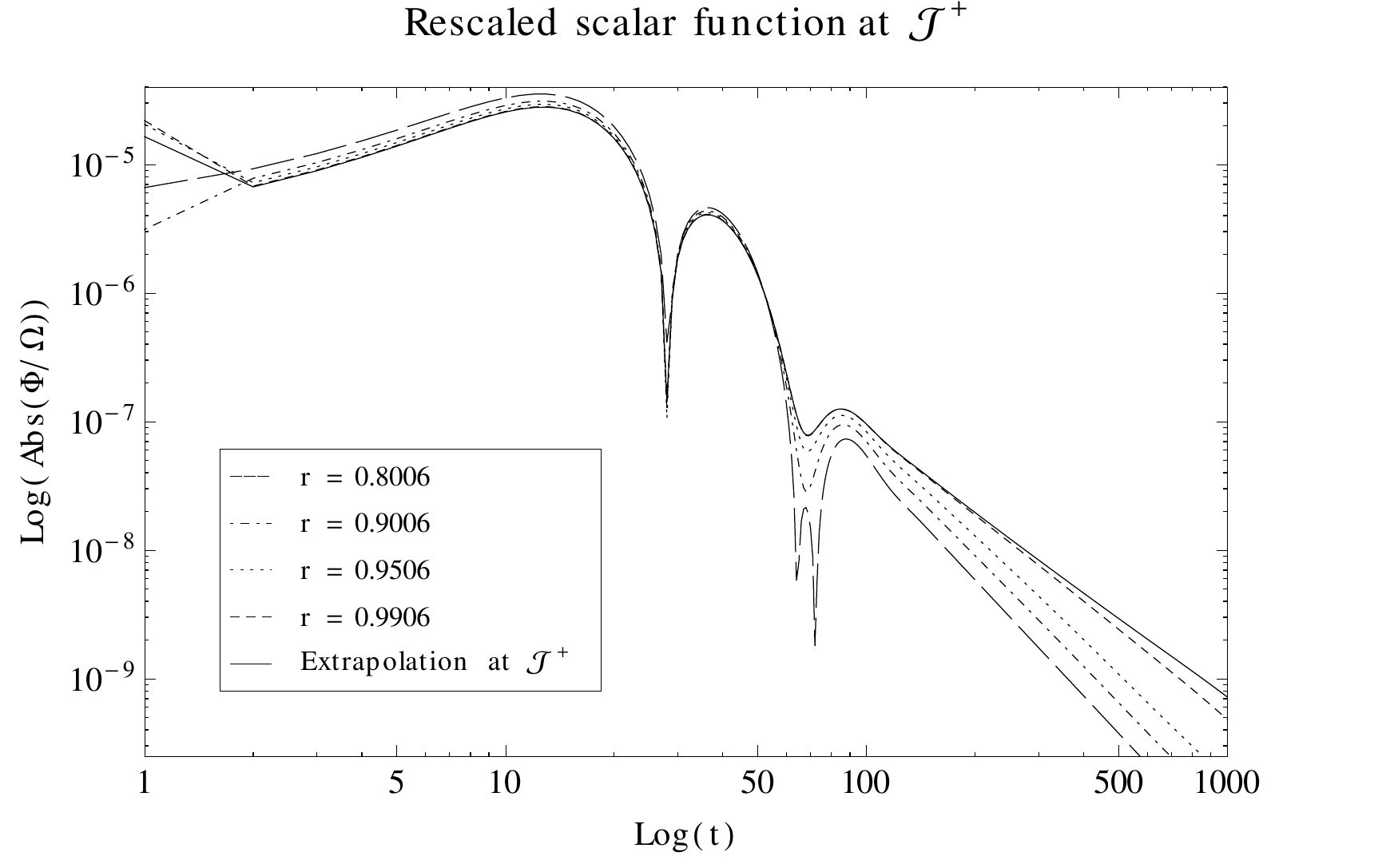}
\caption{Values of the rescaled scalar field at some values of $r$ and extrapolated to $r=r_{\!\!\scri}$ as a function of time. The exponent of the tails is $p\sim-3$ up to $r\sim 0.95$ and $p\sim-2$ at $\scri^+$.}
\end{figure}

We have modified trumpet initial data \eref{fsthyp} with either a perturbation in the initial lapse (gauge waves), or coupled to the massless scalar field with initial data
\begin{equation}
\Phi_0=A\e^{-\frac{(r^2-c^2)^2}{4\sigma^4}}. 
\end{equation}
A scalar field perturbation of a Schwarzschild black hole is expected to decay at late times with a power-law tail of the form \cite{PhysRevD.5.2419}
\begin{equation}
\lim_{t \to +\infty}\Phi(t,r) \propto t^{p}.
\end{equation}
For spherical scalar perturbations the decay rate $p$  has been calculated analytically as $p=-3$ along timelike surfaces \cite{PhysRevD.5.2419} and $p=-2$ along null surfaces ($\scri^+$) \cite{Bonnor66,PhysRevD.49.883}. 
In our simulations we found a value of $p\sim -2.08\pm0.09$ for the scalar field on $\scri^+$. In figure \ref{phitails} we show the behaviour in time of the rescaled scalar field $\Phi/\Omega$ after an initial perturbation of amplitude $10^{-4}$ at some selected values of the radial coordinate, as well as extrapolated to $\scri^+$. 

The convergence results are shown in figure \ref{convphitails}. The values of $\Phi/\Omega$ are extrapolated to $\scri^+$ before calculating the differences presented in the plot. We see good convergence up to $t\approx 100$, when the convergence order starts to decrease due to loss of accuracy. 

We note that some tuning of the gauge conditions is necessary to suppress continuum instabilities in the evolution systems. Some algebraic modifications of the evolution variables which also suppress continuum instabilities are discussed in \cite{Vano-Vinuales:2014koa}.

\section{Conclusions}

We have evolved nonlinear spherical perturbations of a Schwarzschild black hole using a hyperboloidal free evolution approach based on  conformally compactified versions of the generalized BSSN and \CZ{} evolutions systems for the Einstein equations. The present paper extends results we recently obtained for regular initial data \cite{Vano-Vinuales:2014koa} to the presence of a black hole. Our initial data correspond to nonlinear 
perturbations of initial data for black hole trumpet geometries, which we have calculated on the compactified hyperboloidal slices.  
We obtain stable numerical evolutions of the scalar field coupled to the Einstein equations, and show 4th order convergence for the scalar wave signal at $\scri^+$. 
The gauge conditions play a key role in suppressing continuum instabilities in the evolution, and require some degree of tuning. 
Future work will explore more general initial data, gauge conditions, and different versions of the numerical algorithms.

\section*{Acknowledgments}

AV was supported by AP2010-1697, AV and SH were supported by Spanish MINECO grants FPA2010-16495, FPA2013-41042-P and CSD2009-00064, European Union FEDER funds, and the Conselleria d'Economia i Competitivitat del Govern de les Illes Balears. 

\section*{References}
\bibliographystyle{iopart-num_no_url} 
\bibliography{../hypcomp}

\end{document}